\documentclass[aps,pra,reprint,superscriptaddress, longbibliography]{revtex4-2}

\usepackage{siunitx}
\usepackage{amsmath,bm}
\usepackage{amsbsy}
\usepackage{amssymb}
\usepackage{mathptmx, textcomp}
\usepackage{color}
\usepackage{braket}
\usepackage{graphicx}
\usepackage{textcomp}
\usepackage{notes2bib}
\usepackage{textcomp}
\definecolor{bl}{rgb}{0, .1, .6}
\usepackage[colorlinks=true, citecolor = bl, linkcolor = bl, urlcolor=bl, pdfborder={0 0 0}]{hyperref}

\begin{document}
\title{Fast and efficient preparation of 1D  chains and dense cold atomic clouds}

\author{Antoine Glicenstein}
\author{Giovanni Ferioli}
\author{Ludovic Brossard}
\author{Yvan R. P. Sortais}
\affiliation{Universit\'e Paris-Saclay, Institut d'Optique Graduate School, CNRS, 
Laboratoire Charles Fabry, 91127, Palaiseau, France}

\author{Daniel Barredo}
\affiliation{Universit\'e Paris-Saclay, Institut d'Optique Graduate School, CNRS, 
Laboratoire Charles Fabry, 91127, Palaiseau, France}
\affiliation{Nanomaterials and Nanotechnology Research Center (CINN-CSIC),
Universidad de Oviedo (UO), Principado de Asturias, 33940 El Entrego, Spain}

\author{Florence Nogrette}
\author{Igor Ferrier-Barbut}
\author{Antoine Browaeys}
\affiliation{Universit\'e Paris-Saclay, Institut d'Optique Graduate School, CNRS, 
Laboratoire Charles Fabry, 91127, Palaiseau, France}

\begin{abstract}
We report the efficient and fast ($\sim \SI{2}{\hertz}$) preparation of randomly loaded 1D 
chains of individual $^{87}$Rb atoms and of dense atomic clouds trapped in optical 
tweezers using a new experimental platform. This platform is designed for 
the study of both structured and disordered atomic systems in free space. 
It is composed of two high-resolution optical systems perpendicular to each other, 
enhancing observation and manipulation capabilities. 
The setup includes a dynamically 
controllable telescope, which we use to vary the tweezer beam waist.
A D1 $\Lambda$-enhanced gray molasses enhances the loading of the traps from 
a magneto-optical trap.
Using these tools, we prepare chains of up to $\sim 100$ atoms separated by $\sim\SI{1}{\micro\meter}$ 
by retro-reflecting  the tweezer light, hence producing 
a 1D optical lattice with strong transverse confinement. 
Dense atomic clouds with peak densities up to $n_0\sim10^{15}\:\mathrm{at}/\mathrm{cm}^3$
are obtained by compression of an initial cloud. 
This high density results into interatomic distances smaller than 
$\lambda/(2\pi)$ for the D2 optical transitions, making it ideal to study light-induced interactions in dense samples. 
\end{abstract}
\maketitle

\section{Introduction}
The optical response of an ensemble of atoms illuminated by near-resonant light 
can be significantly different from the one of a single atom due to light 
induced dipole-dipole interactions \cite{guerin2016}. They give rise to collective 
behaviors such as modified decay rates or spectral linewidths 
\cite{araujo2016superradiance, roof2016observation, guerin2016subradiance,rui2020subradiant}, 
or resonance shifts \cite{corman2017transmission, Jennewein2018,Glicenstein2020}. 
Recently these effects have drawn an increasing interest, for they can be relevant 
in fundamental optics and have possible applications 
ranging from optical lattice atomic clocks \cite{chang2004,kramer2016,bromley2016} 
to quantum technologies \cite{ostermann2013,plankensteiner2015}.

In order to enhance the collective optical response of an atomic ensemble, 
two different paths can be followed. 
The first one consists in using high-density samples, 
so that the effect of light-induced dipole interactions is large. This requires the preparation 
of atomic clouds with densities $n$ fulfilling $n/k^3\sim 1$ 
where $k=2\pi/\lambda_0$ with $\lambda_0$ the atomic resonance wavelength. 
Fundamental questions arise concerning disordered ensembles, such as the existence of Dicke superradiance in small samples \cite{friedberg1972} or the
saturation of the index of refraction for high densities \cite{Andreoli2021}. In disordered clouds, the field radiated by each emitter acquires a 
random propagation phase that renders difficult the pristine control of interaction effects.  The second path thus consists in spatially structuring 
the cloud at the sub-wavelength scale \cite{bettles2016enhanced,shahmoon2017cooperative}. 
In this way, the interferences can be tailored, making it possible to enhance or suppress 
the effect of dipole interactions. This second route could pave the way to several applications: 
for example, mirrors made by an atomic layer 
\cite{bettles2016enhanced, shahmoon2017cooperative, facchinetti2018interaction}, 
as recently realized using a 2D Mott insulator \cite{rui2020subradiant}, 
controlled transport of excitations \cite{chui2015,needham2019subradiance} and light storage \cite{plankensteiner2015,asenjo2017exponential} or in quantum metrology \cite{ostermann2013,plankensteiner2015,facchinetti2016}. 
The investigation of collective effects in ordered ensembles is also relevant for optical lattice clocks 
\cite{chang2004,kramer2016,campbell2017}, as they could limit their accuracy.

In this paper, we follow the two paths introduced above, relying on a new experimental platform, 
which we describe and characterize. This platform makes it possible to
prepare 1D arrays \cite{karski2009} of \textsuperscript{87}Rb atoms, and
disordered atomic ensembles with peak densities reaching $n_0/k^3\sim 1$. 
This apparatus is an upgrade of our previous experimental setup \cite{Jennewein2018}. 
It consists of two high-resolution optical systems with axes perpendicular to one another in a 
``maltese cross'' geometry similar to \cite{Bruno2019}. These two optical axes used together allow for the simultaneous observation of the 
fluorescence light emitted by the atoms (incoherent response \cite{Pellegrino2014}) 
and the transmission through the cloud (coherent part \cite{Jennewein2016}). One of the axes is used to focus a tight optical dipole trap (tweezer) to confine the atoms. We have placed in the tweezer beam path a telescope made of two lenses with tunable focal length to dynamically control the tweezer waist.  We use this control to prepare chains of atoms with variable length when retro-reflecting the tweezer laser beam, and dense elongated samples after compressing an initially loaded atomic cloud. The loading of the traps from a cloud of laser cooled atoms is enhanced by implementing 
$\Lambda$-enhanced 
gray molasses. 

The paper is organized as follows. 
Section \ref{Sec2:OpticalSetup} describes the optical setup and its alignment, 
the imaging system, and the \emph{OptoTelescope} 
that allows to produce optical tweezers with tunable waists. 
Section \ref{Sec:ChainCharact} presents the realization of a 1D chain with 
controllable length and its characterization. Section \ref{Sec5:GM} 
details the enhancement of the trap loading using 
gray molasses. 
Section \ref{Sec6:compression} introduces a new protocol to prepare dense clouds using the tools described before.

\section{Optical Setup, in-vacuum lenses alignment and imaging system}
\label{Sec2:OpticalSetup} 

\subsection{Optical setup}

\begin{figure}
\includegraphics[width=\linewidth]{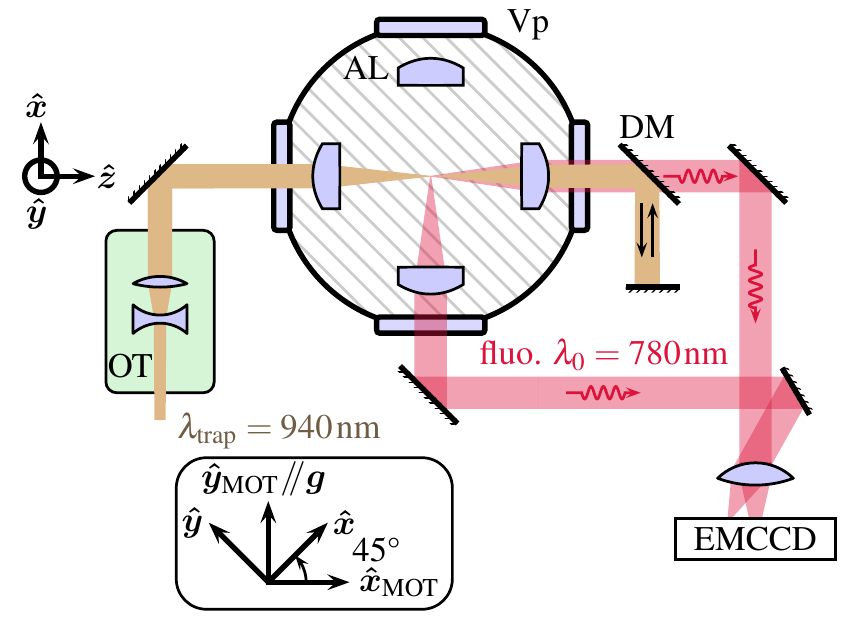} 
\caption{Schematic of the experimental setup. 
Two orthogonal high-resolution (NA=0.44) optical systems based on 
4 in-vacuum aspheric lenses (AL) create an optical dipole trap, which can be 
retroreflected to realize a chain of single atoms 
in a 1D-optical lattice, and collect the scattered light on an 
electron-multiplying CCD (EMCCD) in two perpendicular directions.  
On the tweezer axis, the fluorescence is separated from the trapping 
light using a dichroic mirror (DM). 
The trap radial size is dynamically controlled with the \emph{OptoTelescope} (OT). 
All light enters and exits the vacuum chamber through CF40 viewports (Vp). Insert : The x-axis is rotated by an angle of $\SI{45}{\degree}$ with respect to the plane containing the horizontal beams of the MOT and the z-axis. It is therefore not superimposed to the vertical beam of the MOT, which is in the direction of gravity $\boldsymbol{g}$ .\label{fig1}}
\end{figure}

Trapping individual atoms or preparing dense atomic samples requires 
the waist of the dipole trap beam to be on the order of a few micrometers 
\cite{Schlosser2002,Bourgain2013}. 
This imposes to work with high numerical aperture (NA), diffraction-limited lenses 
\cite{Sortais2007}. As represented in Fig.\,\ref{fig1}, our apparatus is composed 
of four in-vacuum aspheric lenses, forming two orthogonal axes in a
quasi-confocal configuration. The lenses are manufactured by 
Asphericon$^{\mbox{\scriptsize{\textregistered}}}$\footnote{part number AHL25-20-S-U} 
and feature effective $\mathrm{NA}=0.44$. Their working distance ($\SI{15}{\milli\meter}$) is
sufficiently large to allow for large optical access, in particular for the six counter-propagating 
magneto-optical trap (MOT) beams. The plane containing the optical axes of the lenses
makes an angle of $45^\circ$ with respect to the one containing horizontal MOT beams (see Insert Fig.\,~\ref{fig1}): this gives
an extra (vertical) access for the atomic beam entering the trapping region.
This configuration allows the six MOT beams to be orthogonal, 
which facilitates alignment and the overlapping with the dipole trap. 
This also reduces the stray light scattered in the chamber 
and collected by the imaging system.

The conjugated planes has been optimized using an optical design 
software to minimize the aberrations of the two crossed optical systems, at both the 
trapping wavelength $\lambda_{\rm{trap}} = \SI{940}{\nano\meter}$ and 
the \textsuperscript{87}Rb D2 line ($\lambda_0 = \SI{780}{\nano\meter}$), 
the numerical aperture being fixed. Due to the dispersion properties of the glass of the aspheric lenses, the best performances at $\lambda_{\rm{trap}}$ and 
$\lambda_{0}$ are achieved at different focusing positions for initially collimated beams. 
For this reason, we work in a trade-off 
configuration where the optical performances of the lenses are similar 
for the two different wavelengths. More precisely, we impose that 
the wavelength-dependent Strehl ratio ($S$) \cite{BW} is the same at 
$\lambda_{\rm{trap}}$ and $\lambda_{0}$. In our specific case, we calculate 
$S=0.93$,
at a distance $d=+\SI{285}{\micro\meter}$ away from the focal point of a 
lens at $\lambda_0$. For this configuration, we calculate
that the image of an object emitting in vacuum at $\lambda_0$ is located at 
$d_{780}\simeq\SI{1119}{\milli\meter}$ in air (see Fig.\,~\ref{fig2}). This distance 
is used for the alignment procedure of the lenses
described in the next section.

\subsection{In-vacuum lenses alignment}

\begin{figure}
\includegraphics[width=\linewidth]{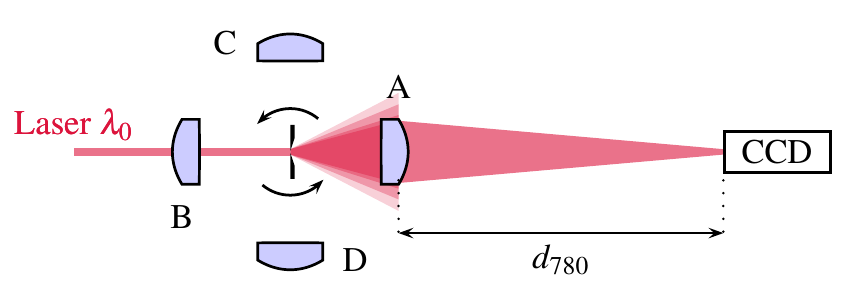} 
\caption{Sketch of the alignment procedure. 
A CCD camera is placed at a fixed position $d_{780}$ 
while we shine a $\lambda_0 = \SI{780}{\nano\meter}$ laser 
beam onto a pinhole acting as a point source for the aspheric lens A. 
By moving the pinhole with respect to lens A, we optimize the Strehl ratio 
on the camera and have access to the best focus of this lens. 
The pinhole is then rotated to face the other lenses. \label{fig2}}
\end{figure}

 The alignment procedure is detailed in \cite{Brossard}. 
 It is experimentally challenging as it involves intersecting two optical axes with a precision much smaller than their field of view ($\SI{\pm 50 }{\micro\meter}$) \footnote{We define the field of view by the region for which the Strehl ratio is larger than 80\% of its peak value}.
 We did the alignment in air, correcting for the difference in index of refraction 
 with respect to vacuum. The barrels holding the aspheric lenses are 
 placed inside a metallic lens holder and separated from it with glass spacers. 
 The lens holder is designed such that the angle formed between the two axes 
 is $\SI{90}{\degree}$ with a tolerance of $\pm  \SI{0.1}{\degree}$. 
 The only degree of freedom for each lens is its on-axis position. 
 It is set by tuning the thickness of the glass spacers with a precision of 
 $\pm \SI{1}{\micro\meter}$. As represented in Fig.\,\ref{fig2}, a CCD camera 
 is first placed at a distance $d_{780}$ away from one lens. A pinhole of diameter 
 $\SI{1 \pm 0.5} {\micro\meter}$ is then mounted on an XYZ translation stage 
 and a rotation stage and placed inside the lens holder. This pinhole is not small 
 enough to be considered as a point source when illuminated by a laser beam 
 at $\lambda_0$. We have taken its finite size into account for the characterization 
 of the performance of the lenses \cite{Brossard}. 
 The pinhole is first moved parallel to the lens axis to minimize the size of its image on the CCD. 
 Once the pinhole is in the targeted object plane, we move it in the transverse 
 plane to maximize the Strehl ratio, thus placing it on the lens optical axis. 
 The pinhole is then rotated by $90^\circ$ to face another lens. This procedure is performed for 
 each lens and by keeping track of the pinhole motion, we obtain a mapping of the best foci. 
 Finally, the  spacers thickness is adjusted to bring all the foci at the 
 same point. After the procedure, we obtain a satisfying alignment of the 
 lenses 
 and the optical axes cross with a residual offset  
 $\lesssim\SI{5}{\micro\meter}$, smaller than the field of view of the lenses.
 
\subsection{Imaging system}

The atoms held in the dipole trap are imaged with the two high-resolution axes (Fig.\,\ref{fig1}), with a
diffraction-limited resolution of $1.22\lambda_0/(2NA)\simeq \SI{1}{\micro\meter}$. 
Along the trapping axis $\boldsymbol{\hat{z}}$, the fluorescence or the transmitted light is 
separated from the trap light using a dichroic mirror and interferometric filters, 
and is collected by an electron-multiplying CCD (EMCCD) with pixel size 
$\SI{16}{\micro\meter}\times\SI{16}{\micro\meter}$ \footnote{Andor iXon Ultra 897}. 
The magnification of the imaging system along this axis is $6.4$, leading to an effective pixel size 
in the object plane of $\SI{2.5}{\micro \meter}$:  this allows focusing the light emitted by a 
single trapped atom onto a single pixel, maximizing the signal-to-noise ratio, 
albeit at the cost of a lowered resolution with respect to the diffraction limit. 
The fluorescence emitted along the $\boldsymbol{\hat{x}}$-axis is collected on 
the same camera, allowing for the observation of the atoms in two orthogonal directions 
in a single image. The magnification on the transverse axis is $\sim 16$, leading to an 
effective pixel size of $\SI{1}{\micro\meter}$ in the object plane. Both resolutions were verified using 
calibrated pinholes in planes conjugate to the atoms plane. The magnification was confirmed 
by measuring simultaneously the displacement of trapped atoms on both axes when moving 
the trapping beam by a known distance. 
The estimated collection efficiency of both imaging systems is $\sim4\%$, taking into account 
the collection of the aspheric lens (5\%), 
the transmission of the optical elements (90\%) 
and the camera quantum efficiency (85\% at $\lambda_0 = \SI{780}{\nano\meter}$). This value is confirmed by the measurement of the fluorescence at saturation of a single \textsuperscript{87}Rb atom in a tight dipole trap. 
As detailed below, we use this atom as a probe to characterize the trap (size and depth), as was done in \cite{Sortais2007}.

\subsection{The \emph{OptoTelescope}}

Our apparatus includes a telescope with  
tunable magnification, which we name here \emph{OptoTelescope} (OT). 
This telescope is composed of a pair of 1 inch lenses with voltage-controlled focal lengths, manufactured 
by OptoTune$^{\mbox{\scriptsize{\textregistered}}}$ \footnote{part number EL-10-30 Series}, 
and placed in an afocal configuration. Tunable lenses allow for the dynamical manipulation of dipole traps \cite{leonard2014}. Here, using the OT, 
we dynamically change the size of the trapping beam before the aspherical lens
and thus the optical tweezer waist. 
To limit aberrations from the OT, 
we use a beam diameter of 
$\simeq\SI{1}{\milli\meter}$ at its entrance.
Also, we minimize gravity-induced aberrations by positioning the optical axis of the lenses vertically.  
In order to achieve 
small waists on the atoms, the beam after the OT is magnified by a $\times 4$ 
telescope before being focused by the aspherical lens. 
The OT is designed for a magnification ranging from 1 to 3.5. 

\begin{figure}
\includegraphics[width=\linewidth]{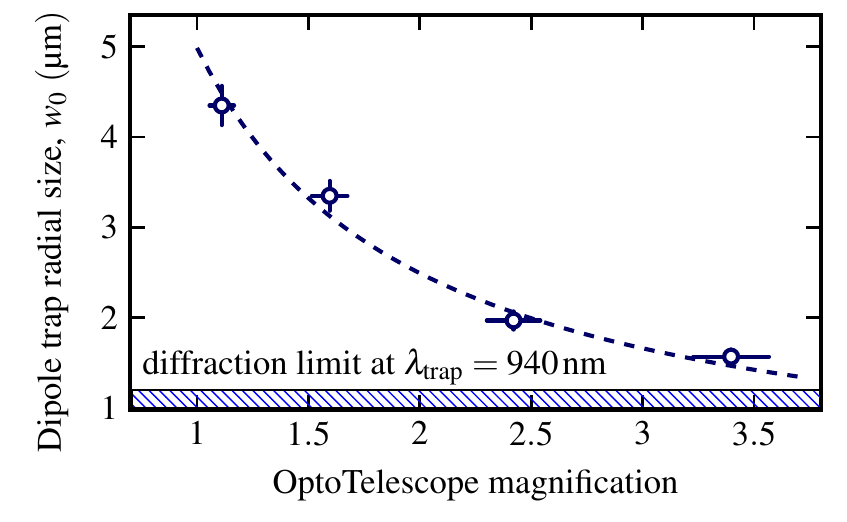} 
\caption{Dipole trap waist at $1/e^2$ as a function of the OT magnification. 
The diffraction limit $1.22\lambda_{\rm trap}/2NA\simeq\SI{1.15}{\micro\meter}$ is indicated as the smallest trap achievable with the apparatus. 
The dashed line corresponds to the expected size. \label{FigOptoWaist}}
\end{figure}

We characterized  the \emph{OptoTelescope}  by performing \emph{in situ}  
measurements on a single atom trapped in the tweezer. 
For a given magnification, the waist 
of the trap $w_0$ is measured as follows. For a fixed power $P$, the peak 
intensity and thus the light-shift induced by the trap (proportional to the trap depth $U$) 
are obtained by using a push-out beam expelling the atom from the trap. 
The light shift is measured from the detuning of this beam for which the push-out effect is the largest, recorded for various trap depths. 
The trap waist is then extracted using $U \propto P/w_0^2$. 
The results were checked by independent measurements of
the oscillation frequencies of individual trapped atoms \cite{Sortais2007}. 
We are able to dynamically change the size of the trap between about $\SI{1.6}{\micro\meter}$ and 
$\SI{4.3}{\micro\meter}$, in agreement with the theoretical values calculated 
using gaussian optics, as shown in Fig.\,\ref{FigOptoWaist}.

\section{\label{Sec:ChainCharact} Realization of  a chain of atoms with controllable length}

In this section, we present the preparation and characterization of
one-dimensional atomic chains of cold Rb atoms, using the tools described in the previous section.

\begin{figure}
\includegraphics[width=\linewidth]{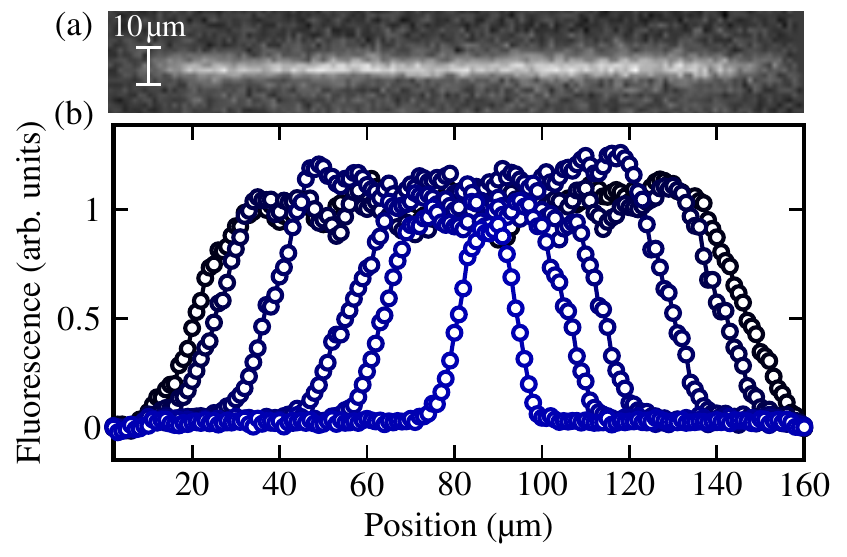} 
\caption{(a) Averaged image of the fluorescence collected by the transverse imaging axis. 
(b) Cuts of the fluorescence along the chain for various chain lengths. \label{FigChainLength}}
\end{figure}

As represented in Fig.\,\ref{fig1}, we produce the chain by retro-reflecting the tweezer beam
using the second aspherical lenses placed on the same axis, 
thus forming a 1D optical lattice with an 
inter-site spacing $\lambda_{\rm{trap}}/2 = \SI{470}{\nano\meter}$ \footnote{The retroreflected beam has an intensity  
reduced by half because of the transmission through all the optical elements.}. The small beam waist of the
tweezer ensures a tight transverse confinement.  
This 1D array in then loaded from the MOT with a filling fraction 
averaged along the chain of $\simeq 0.25$. We will show in the next section that the loading 
can be improved up to $\sim0.5$ using  gray molasses. We collect the fluorescence emitted by the chain 
in the transverse direction under a $\SI{20}{\milli\second}$ excitation by the MOT beams. 
A typical example of the atomic 
chain  is shown in Fig.\,\ref{FigChainLength}(a)
(the resolution being about twice the inter-trap separation, 
we do not resolve individual sites).
The length of the atomic chain is given by the range around the focal 
point where the beam intensity is high enough to trap atoms, which is set by the Rayleigh distance 
$z_{\rm{R}}=\pi w_0^2/\lambda_{\rm{trap}}$. 
Experimentally, we realize atomic chains 
with different lengths (and atom number) by tuning the waist of the trapping beam
using the OT. 
As changing the waist also modifies the trap depth, we adapt its power 
to keep the depth at the center of the chain at $\sim\SI{1}{\milli\kelvin}$. 
In Fig.\,\ref{FigChainLength}(b) we present cuts of the fluorescence along the chain for 
various lengths. Our longest chains have lengths of 
$\sim \SI{100}{\micro\meter}$ (hence $\sim200$ sites).

To characterize the chain, we measure the local transverse and longitudinal trapping frequencies 
$\omega_r$ and $\omega_z$ along the chain axis. To do so,  we rely on parametric heating 
by modulating the intensity of the trapping beam at a given frequency, inducing losses 
at $2\omega_r$ or $2\omega_z$. Since the trap depth varies along the chain, 
the oscillation frequencies depend on the position, 
and so do the resonant frequencies of the parametric heating. 
Experimentally, we first load a chain from the MOT and take a first reference fluorescence image.
The trap beam power is then set at a value of $\SI{140}{\milli\watt}$ while, for this measurement, 
the waist is set to $\SI{3.3}{\micro\meter}$. 
%
With these parameters we expect $\omega_z \simeq 2\pi\times\SI{1}{\mega\hertz}$ 
and $\omega_r \simeq 2\pi\times\SI{70}{\kilo\hertz}$ at the center of the chain. 
The beam intensity is then modulated with a relative amplitude of 5\% 
during $\SI{100}{\milli\second}$ using an arbitrary waveform generator.
A second fluorescence image of the chain is then taken and compared to the reference 
image to evaluate the atom losses. This sequence is repeated 50 times 
to average over the chain filling.

\begin{figure}[t!]
\includegraphics[width=\linewidth]{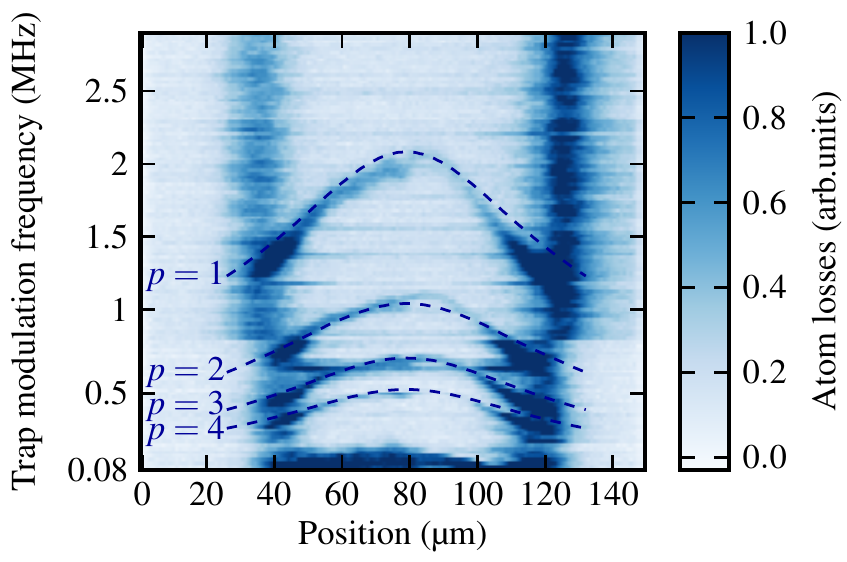} 
\caption{ Atom losses as a function of the position in the chain 
and the modulation frequency of the trapping beam. 
The dashed lines are the calculated axial oscillation frequencies. 
The multiple resonances correspond to $2\omega_z/p$ with $p$ integer. 
\label{OscillationsFreq}}
\end{figure}

Figure \ref{OscillationsFreq} shows the atom losses due to the axial excitation. 
The resonance frequencies extracted with this method are in good agreement with 
the calculated oscillation frequencies (dashed lines), confirming the expected value 
of the waist. The different dashed lines reported in  Fig.\,\ref{OscillationsFreq}, 
are given by $2\omega_z/p$ with $p$ integer. We observe losses at these 
frequencies since the amplitude modulation is not perfectly sinusoidal and 
thus contains tones at multiples $p$ of the driving frequency. 
We also observe
losses on the chain edges where the trap is the shallowest: these are due 
to the reference imaging light expelling atoms from the shallow traps, which are thus 
not recovered in the second fluorescence image. The same experiment was done 
for radial oscillation frequencies, obtaining also in this case a good agreement 
between the measured trapping frequencies and the predicted ones.

\section{\label{Sec5:GM}Optimization of the loading using $\Lambda$-enhanced gray molasses}

Gray molasses (GM) are commonly used to achieve 
sub-Doppler cooling of atoms using dark states \cite{grynberg1994,boiron1995,boiron1996,esslinger1996,fernandes2012}. 
The use of GM in a tight optical tweezer offers two interesting prospects.
First, the low photon scattering rate 
in dark states  reduces light-induced collisions. This yields a higher density of the atomic cloud the tweezer
is loaded from, and hence a larger number of atoms  in the tweezer. 
Second, their blue detuning with respect to the atomic frequency 
should permit to tailor light-induced collisions to selectively remove 
a single atom out of a pair, resulting into exactly one atom per trap 
with high probability \cite{grunzweig2010,Brown2019}.

We first consider the loading of a single atom in a small (non-retroreflected) tweezer, 
and apply $\Lambda$-enhanced gray molasses \cite{Grier2013} on the \textsuperscript{87}Rb D2 line 
($\lambda_0=\SI{780}{\nano\meter}$) \cite{Rosi2018}. 
The cooling beam is blue-detuned from the $(5S_{1/2},F=2)$ to $(5P_{3/2},F'=2)$ transition 
and superimposed with the six MOT beams with intensity
 $I \sim I_{\mathrm{sat}} = \SI{1.67}{\milli\watt\per\square\centi\meter}$ per beam. 
 The coherent repumper is created 
from the same laser using an electro-optical modulator with frequency equal to the 
ground state hyperfine splitting $\nu = \SI{6834.68}{\mega\hertz}$. 
The intensity of the repumper is $I \sim I_{\mathrm{sat}}/10$ per beam, 
given by the sideband amplitude. 
Since gray molasses rely on the blue detuning of the cooling lasers, 
the optimal detuning will depend on the light-shift induced by the tweezer beam. 
After the MOT beams are switched off, we study the loading of a single atom from the GM
into the tweezer (waist $w_0 = \SI{1.6}{\micro\meter}$) 
varying the detuning of the GM and the trap depth. For each set of parameters, 
we record the loading probability and the atom temperature, using a release and recapture method 
\citep{Sortais2007,Tuchendler2008}.
We have found  that using the GM on the D2 line does result into individual atoms in the tweezer being
colder than when loaded directly from the MOT 
($\sim\SI{20}{\micro\kelvin}$ instead of $\SI{80}{\micro\kelvin}$), 
and for a much broader range of the tweezer depth. 
Also, when  loading directly from the MOT, the atoms can be captured in traps with 
depth $U/k_{\rm{B}} \sim \SI{1}{\milli\kelvin}$ 
while applying the GM stage allows trapping for depth down to $U/k_{\rm{B}} \sim \SI{200}{\micro\kelvin}$. 
Furthermore, we  observe that the GM detuning does not significantly change the temperature 
or the loading over a wide range of parameters for detunings between $50$ and $\SI{120}{\mega\hertz}$ 
above the transition and depths $U/k_{\rm{B}}$ between $\SI{200}{\micro\kelvin}$ and $\SI{1}{\milli\kelvin}$. 
For larger trap depths and small detunings, the GM frequency becomes resonant with the 
$ (5S_{1/2},F=2)$ to $(5P_{3/2},F'=2)$ transition, resulting in heating of the atom.
However, while we observe efficient cooling when applying the GM, we have not found loading 
probabilities significantly higher than 50\% in a single tweezer, or 25\% in chains of traps (retroreflected tweezers), 
similar to what we achieved with the MOT. 
This  might be due to the fact that the blue-detuned beam on the $(5S_{1/2},F=2)$ to $(5P_{3/2},F'=2)$ 
transition is  detuned to the red of the $(5S_{1/2},F=2)$ to $(5P_{3/2},F'=3)$ transition 
($\SI{267}{\mega\hertz}$ higher in frequency), causing light-induced collisions, 
which may limit the loading.

\begin{figure}
\includegraphics[width=\linewidth]{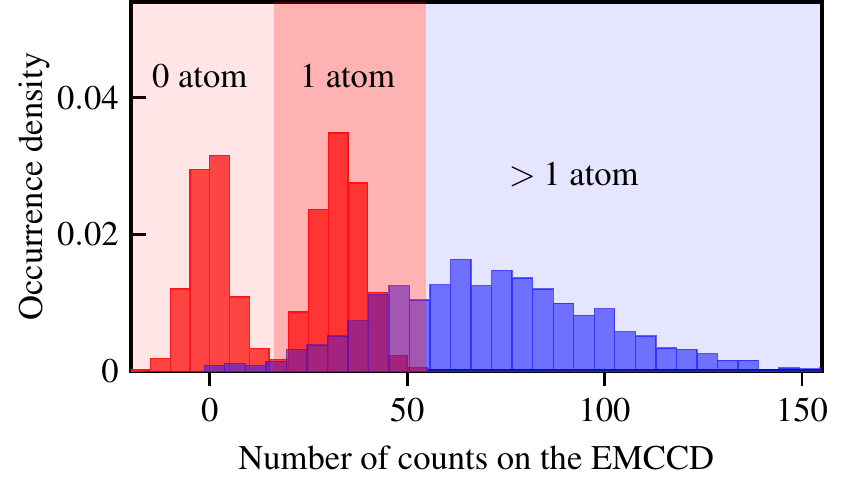} 
\caption{ Histogram of the collected fluorescence 
of atoms in a  trap loaded 
with the GM (blue), in comparison with a trap loaded with a single atom (red). The fluorescence is induced in both cases by $\SI{20}{\milli\second}$ of MOT beams with detuning $3\Gamma$, where $\Gamma$ is the natural linewidth of the \textsuperscript{87}Rb D2 line. A background image, without atom, has been subtracted.
\label{histograms}}
\end{figure}

To circumvent this issue, we have thus implemented gray molasses on the D1 line [$(5S_{1/2},F=2)$ to $(5P_{1/2},F'=2)$ transition]. In the single non-retroreflected tweezer, after optimization, we were not able to obtain individual atoms 
with a probability significantly higher than 50\%, whatever the detuning. This is in contrast to what was reported using a blue-detuned beam \cite{grunzweig2010} or  GM on the D1 line \cite{Brown2019}.
To explain this observation, we compare the volume of our tweezer 
to the one used in Ref.~\cite{Brown2019} and estimate ours to be a factor of $>10$ larger. 
Thus our collision rate is reduced by this 
factor and the time for blue-detuned light-induced collisions 
to induce selective losses and leave a single atom in the trap 
should be much longer than experimentally achievable timescales.
We thus infer that more than one atom are left inside the trap. 
To confirm this, we compare the result of loading via the GM, with the direct loading from the MOT. In one case, we load directly the trap from the MOT: the collisional blockade mechanism operates \cite{Schlosser2002,Sortais2007} and when sending near resonant light for imaging, 
we observe two clear fluorescence levels corresponding to either 1 or 0 
atom in the trap. In the other case, we apply a 
$\SI{200}{\milli\second}$-long GM to load the trap and then image the atoms as before. 
Under this condition, we record a broad 
fluorescence histogram, as shown in  Fig.\,~\ref{histograms}. We explain it
by the fact that the initial atom number is large. However the imaging light 
induces strong losses removing the atoms during the imaging time thus preventing us from counting 
precisely the \emph{in-situ} atom number.\par
Finally, we have used D1 gray molasses to improve the loading of the atom chain. 
We are now able to load a chain of traps with a 50\% probability. This is likely due to the fact that on average there are more than one atom per site following the gray molasses loading. The application of the MOT light for imaging then induces strong light-induced collisions, leaving either 0 or 1 atom. Further investigations will be necessary to unravel the loading mechanism of this chain of closely-spaced traps by D1 $\lambda$-enhanced gray molasses. We have also found that the loading using GM is
more stable than the direct loading from the 
MOT in terms of daily fluctuations. 


\section{\label{Sec6:compression} Preparation of dense atomic clouds }

As mentioned in the introduction, one of the motivations for our 
new set-up is the  study of light scattering in dense ensembles. 
We present here a loading protocol based on 
the new tools of the setup that allows preparing dense enough samples. 
The main idea is to load as many atoms as possible into a large single-beam dipole 
trap using GM on the D1 line, and compress the cloud by dynamically reducing the beam waist \cite{Kinoshita2005} using the \emph{OptoTelescope}. 
 
We start from a 3D-MOT, which is  compressed in $\SI{15}{\milli\second}$ 
by red-detuning the MOT beams from -$3\Gamma$ to -$5\Gamma$. We then decrease the magnetic field gradient by 50\%. 
The MOT beams are then switched off and the GM is applied for 
$\SI{200}{\milli\second}$, with the dipole trap on. 
At this stage, the trap depth is $U/k_{\rm B}\simeq\SI{4.2}{\milli\kelvin}$ 
and the waist is $w_0\simeq\SI{2.5}{\micro\meter}~$\footnote{We do not use a larger 
waist because larger initial waists are accompanied by a large axial displacement 
of the focal position  when compressing a cloud, inducing significant heating and atom losses.}. 
In this starting configuration, we trap up to 6000 atoms at a temperature of 
$\SI{625}{\micro\kelvin}$ yielding a peak density $n_0\approx \SI{1.6 e14}{at\per\cubic{\centi\meter}}$. The use of GM is the key ingredient here that allows for the loading of this large number of atoms. 
The cloud has an aspect ratio of about 12 along the trapping axis. 
The atom number is evaluated from the fluorescence 
collected during the  illumination of the cloud with a $\SI{10}{\micro\second}$-pulse of resonant light 
and dividing the signal by the same quantity measured with a single atom. 
To avoid effects caused by light-induced interactions, the imaging pulse in sent 
after a time-of-flight of $\SI{10}{\micro\second}$ during which the density 
drops by about an order of magnitude 
\footnote{We have verified that the atom number obtained at this density is the same than the one obtained with a 
5 times longer time-of-flight, after which the cloud is dilute ($n_0/k^3 \ll 1$).}. 
The temperature is measured by fitting the cloud size for a variable time-of-flight.

The trap is then compressed to a waist $w_0 = \SI{1.8}{\micro\meter}$ by 
changing the magnification of the \emph{OptoTelescope} in $\SI{30}{\milli\second}$, 
keeping the power constant. Next, the trap depth is  increased in $\SI{10}{\milli\second}$ 
up to $\SI{7.6}{\milli\kelvin}$. The duration of the compression has been optimized to 
be short enough to minimize three-body losses but long enough compared to the 
response time of the OT lenses ($\SI{2.5}{\milli\second}$). 
At this stage, we obtain a cloud of about 2500 
atoms in the trap 
at a temperature of $\SI{700}{\micro\kelvin}$, which corresponds to a cloud peak density 
$n_0\sim 10^{15}\:\mathrm{at}/\mathrm{cm}^3$ or equivalently to 
$n_0/k^3= 1.7 \pm 0.3$. This density is three times larger than the one obtained in clouds of $\sim 500$ atoms \citep{Bourgain2013,Pellegrino2014} with our previous apparatus which relied on a first, large dipole trap acting as a reservoir to load a second small tweezer. 

\begin{figure}
\includegraphics[width=\linewidth]{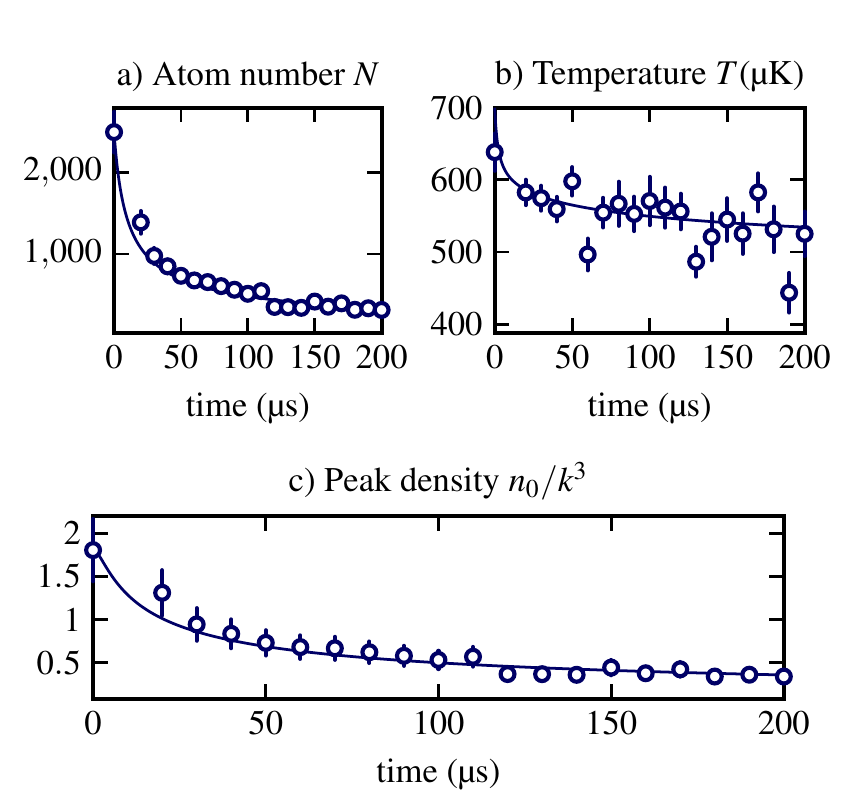} 
\caption{Time evolution in the final trap of atom number (a) and temperature (b).
In (a) and (b), the solid lines correspond to the solutions of \eqref{decN} and 
\eqref{decT} with $L_3$ as single fit parameter. 
(c) Peak density $n_0$ in the trap, deduced from (a) and (b). \label{DensityFigure}}
\end{figure} 

Such a high density results in large 3-body losses and high elastic collision 
rates. To characterize them and confirm the extracted value of the density, 
we study its dynamics. To do so,
we have measured the cloud atom number $N$ and temperature $T$ as a function of the time after the 
end of the compression. The results are shown in Fig.\,\ref{DensityFigure}(a). 
The temporal evolution of $N$ and $T$ 
is described by the following system of coupled equations that takes into account 
of 2- and 3-body losses \cite{Bourgain2013,luiten1996,eismann2016}:
\begin{eqnarray}
\frac{dN}{dt}&=& -\gamma_3 \frac{N^3}{T^5} - \gamma_2 \left( \sigma(T),T \right) \frac{N^2}{T} \label{decN}\\
\frac{dT}{dt}&=& \frac{T}{3}\left[ \frac{5}{3} \gamma_3 \frac{N^2}{T^5} - \tilde{\gamma}_2 \left( \sigma(T),T \right) \frac{N}{T}\right]\label{decT}
\end{eqnarray}
where the parameter $\gamma_3$ depends on the trap geometry and 
is proportional to the 3-body losses coefficient $L_3$. 
The coefficients $\gamma_2$ and $\tilde{\gamma}_2$ depend on the temperature, the trap geometry 
and on the two-body elastic cross-section $ \sigma(T)$, whose temperature dependence takes into account the 
$d$-wave resonance at $\SI{350}{\micro\kelvin}$. We interpolate the data of \cite{Buggle2004} to find a functional form of $\sigma(T)$. 
We fit the decay of the atom number with the solution of Eq. \eqref{decN}, 
leaving solely $L_3$ as a fit parameter. We obtain 
$L_3 = \SI{4 \pm 1 e-28}{\centi\meter^6\per\second}$. 
This value is larger 
\footnote{Taking into account the reduction of losses in a Bose Eintein condensate by a factor of 6.}
than those found in the literature \cite{Soding1999a,Burt2008}.
Note that there exists no prediction for the effect of the $d$-wave resonance on 3-body losses, 
which could enhance $L_3$ at $T=\SI{650}{\micro\kelvin}$. 
We thus do not expect to find the literature  values, which were measured deep in the
$s$-wave regime. 
We also compare the model prediction of the temperature evolution to the data 
[see Fig.\,\ref{DensityFigure}(b)], and find a very good agreement. 
The temperature is almost constant, which justifies the assumption of a 
temperature-independent $L_3$ (and hence $\gamma_3$) in the model.
Combining the measurements of the atom number and of the temperature, 
we calculate the cloud density. Its evolution is shown in Fig.\,\ref{DensityFigure}(c). 

Our experiment is therefore able to efficiently produce microscopic clouds containing up to 
a few thousand atoms at densities $n_0\sim k^3$.   
This corresponds to the regime where the atoms become strongly 
coupled by light-induced resonant dipole-dipole interactions 
(scaling as $\hbar\Gamma/(kr)^\alpha$ with $\alpha=1,2,3$). 
Moreover the repetition rate of the experiment is high:
about $\SI{2}{\hertz}$, limited by the MOT loading. Thanks to this, fast data acquisition is possible, which has allowed us to observe and control subradiance in the time domain \cite{ferioli2020}. It is in addition a strong asset when measuring, e.g., intensity correlations of the light emitted by the atomic 
ensemble.

\section{Conclusion}

We have built an experimental setup that is well-suited for the study of 
light scattering in cold atom ensembles either in an ordered or disordered configuration. 
Our platform combines two high-resolution optical systems
perpendicular to each other, an optical tweezer with a dynamically 
tunable waist and gray molasses on the D1 line. 
By retroreflecting the optical tweezer we create 
an optical lattice of controllable length,  
allowing for the preparation of atomic arrays with an average interatomic distance 
$1.2\,\lambda_0$. We recently used this feature to investigate a collective enhancement 
of light-induced interactions in 1D arrays \cite{sutherland2016collective,Glicenstein2020}.  
The same strategy can be applied with an optical lattice of shorter 
wavelength (e.g. combining a repulsive optical lattice at $\SI{532}{nm}$
with the infrared tweezer for confinement). This would increase 
collective effects even further, enabling the observation of subradiant 
modes in ordered arrays \cite{bettles2016cooperative,asenjo2017exponential}. 
Furthermore, we presented a protocol for preparing
dense clouds in a tightly focused optical tweezer  
that exploits the dynamical tunability of the OT. 
In this way we create clouds with a peak density larger than 
$k^3$ at a rate $>\SI{2}{\hertz}$. The short inter-atomic 
distances reached in this configuration also offers interesting prospects for
investigations of superradiance in small ensembles and subradiance as we recently reported in \cite{ferioli2020}, as well as the study of fundamental questions 
such as the saturation of the refractive index of dense media \cite{Andreoli2021}.

\begin{acknowledgements}
	We thank Brandon Grinkenmeyer for early work in the construction of the apparatus. This project has received
funding from the European Union’s Horizon 2020 research
and innovation program under Grant Agreement
No. 817482 (PASQuanS) and by the R\'egion \^Ile-de-France
in the framework of DIM SIRTEQ (project DSHAPE) and DIM Nano-K (project ECONOMIQUE). 
A.\,G.~is supported by the D\'el\'egation G\'en\'erale de l’Armement Fellowship No.\,2018.60.0027.
\end{acknowledgements}

\bibliography{setup}
\end{document}